# Generation of incoherent light from a laser diode based on the injection of an emission from a superluminescent diode

**Akifumi Takamizawa\*, Shinya Yanagimachi, and Takeshi Ikegami**

*National Metrology Institute of Japan (NMIJ), National Institute of Advanced Industrial Science and Technology, Tsukuba, Ibaraki 305-8563, Japan*

*\*Corresponding author: akifumi.takamizawa@aist.go.jp*

In this study, incoherent light with a spectral linewidth of 7 nm and 140 mW of power was generated from a laser diode into which incoherent light emitted from a superluminescent diode was injected with 2.7 mW of power. The spectral linewidth of the light from the laser diode was broadened to 12 nm when the diode's output power was reduced to 15 mW. In the process of transformation from single-mode laser light to incoherent light with a broad spectrum by increasing injection-light power, multimode laser oscillation and a noisy spectrum were found in the light from the laser diode. This optical system can be used not only for amplification of incoherent light but also as a coherence-convertible light source.

## 1. Introduction

Incoherent light with a broad spectrum of several tens of nanometers is used for applications based on low coherence interferometry such as optical gyroscopes [1-3] and optical coherence tomography [4-6]. In the latter, while a broader spectrum results in higher axial resolution, light power is also important for better sensitivity [5]. Superluminescent diodes (SLDs) have recently been developed to overcome the difficulty of creating an incoherent light source with high output power [7-10].

As light emitted from an SLD has high spatial coherence due to its micrometer-size output area in contrast to its temporal coherence, a taper amplifier with a tiny input area can be employed to enhance the power of light from an SLD. Indeed, a light source in which an SLD and a taper amplifier are integrated on a chip has been developed [10]. The production of output power beyond 1 W from an SLD with a taper active region has also been reported [7].

Meanwhile, an injection locking method involving a master light incident to a slave laser cavity has been used to phase-lock the slave laser light to the master [11-15]. In many applications, coherent light has been used as the master; for example, injection locking of a high output-power laser diode (LD) to light emitted from an external cavity diode laser or an optical frequency comb has been employed to make a tunable and high-intensity laser source with a narrow linewidth [11-13].

Here, let us consider a case in which the master light is provided from an SLD. Under phase locking, the slave light will have the same spectrum as the master light; that is, it will become incoherent. However, the injection lock is unstable basically when the frequency of the master light is far from the resonant frequency of the slave laser's cavity. Hence, it may be considered that the spectrum of the slave light has many narrow peaks arranged with the intervals of a free spectral range (FSR).

Nonetheless, successful injection locking of the slave laser to the incoherent master light will provide an alternative to the use of a taper amplifier for enhancing the power of incoherent light. The method is expected to be particularly important in optical wavelength ranges that are outside the active range of taper amplifiers.

Additionally, injection locking can be easily switched off by interrupting the injection of the master light with a mechanical shutter or similar. This approach has been employed in atomic fountain clocks, in which light shift (i.e., the frequency shift caused by residual laser light with a frequency close to that of the atomic resonance) is suppressed by stopping the injection of a master light tuned to the vicinity of the atomic resonant frequency [14]. In the case of injection locking for incoherent light, the slave light is switched between incoherent light with a broad spectrum and laser light with a narrow linewidth. Hence, the optical system works as a coherence-convertible light source.

This paper describes an experiment in which incoherent light emitted from an SLD was injected into an LD. Although the LD was not completely phase-locked, the laser light from the diode, whose power exceeded 100 mW, unexpectedly varied into incoherent light with a spectral linewidth of around 10 nm at a master-light power of a few mW. Incoherent light from the LD also reverted to laser light when master-light injection was interrupted.

Section 2 describes the experimental setup and the typical spectra of slave light for various levels of master-light power. The linewidth and other characteristics of the spectra are then evaluated for various levels of master- and slave-light power. Section 3 outlines how higher power and a broader spectrum can be achieved in the injection method, and explains why further advanced theoretical analysis based on conventional theoretical estimation is necessary to quantitatively explain the experimental results. Finally, the conclusion is given in Section 4.

## 2. Experiment

### A. Experimental Setup

The experimental setup is schematically shown in Fig. 1. The master light emitted from the SLD had a central wavelength of 850 nm and a spectral linewidth of 50 nm (full width at half maximum; FWHM) as shown in Fig. 2. The master light passed through an optical isolator (OI1) and was coupled to a single-mode fiber to improve the beam pattern. Slave light radiated from the Fabry-Perot-type LD also passed through another optical isolator (OI2) with a transmittance of 87% and entered an optical spectrum analyzer (OSA) to allow measurement of its spectrum. Here, the cw laser light from the LD without injection had a single longitudinal mode and a wavelength of around 855 nm as shown in Fig. 3(a). The FSR of the LD was 0.8 nm. The master light output from the fiber was collimated and shone onto the polarization beam splitter (PBS) inside the OI2 from the side. The s-polarization component of the master light was reflected from the PBS and injected into the LD after the polarization was rotated using the Faraday rotator inside the OI2 for agreement between the master and slave lights in terms of polarization. Here, small reflection of the slave light from the PBS (not shown in Fig. 1) was coupled to the fiber, and the coupling efficiency was maximized. In this way, the alignment of the master light to the LD was simultaneously optimized.

The values $P_m$ and $P_s$ used here indicate the master- and slave-light power between the LD and the OI2, respectively, as depicted in Fig. 1. The maximum value of $P_m$ was 2.7 mW. To maintain the spectrum of the master light, its power was varied by rotating the half-wavelength plate in front of the PBS in the OI2 rather than by changing the current injected into the SLD. The slave-light power $P_s$ was set to 15, 51 or 150 mW by adjusting the current injected into the LD.

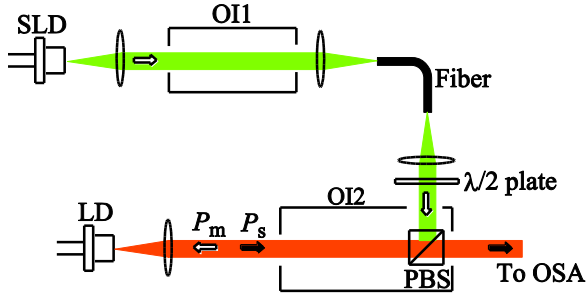

Fig. 1. Schematic representation of the experimental setup. The black and white arrows indicate the propagation directions of the slave and master lights, respectively. The path of the master light was superimposed onto that of the slave light using an OI2 optical isolator. The master- and slave-light power between the LD and the OI2 are indicated by $P_m$ and $P_s$, respectively.

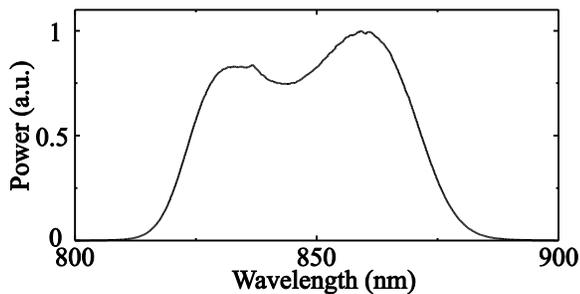

Fig. 2. Spectrum of emission from the SLD.

### B. Results

The black lines in Fig. 3 show the spectra of the slave light with power $P_s$ = 51 mW and master-light power $P_m$ = 0, 0.3, 0.8 and 2.7 mW. As seen in Fig. 3(a), a narrow peak caused by single-mode oscillation appeared in the spectrum when no master light was injected. In contrast, when a weak master light was injected, a broad spectrum appeared together with multimode laser oscillation as shown in Fig. 3 (b). As the power of the master light increased, the laser modes became unclear and the spectrum became noisy as shown in Fig. 3(c). The noisy component also became smaller with increased master-light power as shown in Fig. 3 (d). Overall, as the master-light power increased, the linewidth of the spectrum's broad component became larger while the peak power of the spectrum decreased. Here, the slave-light power was independent of the master-light power, including that in the case with no master light.

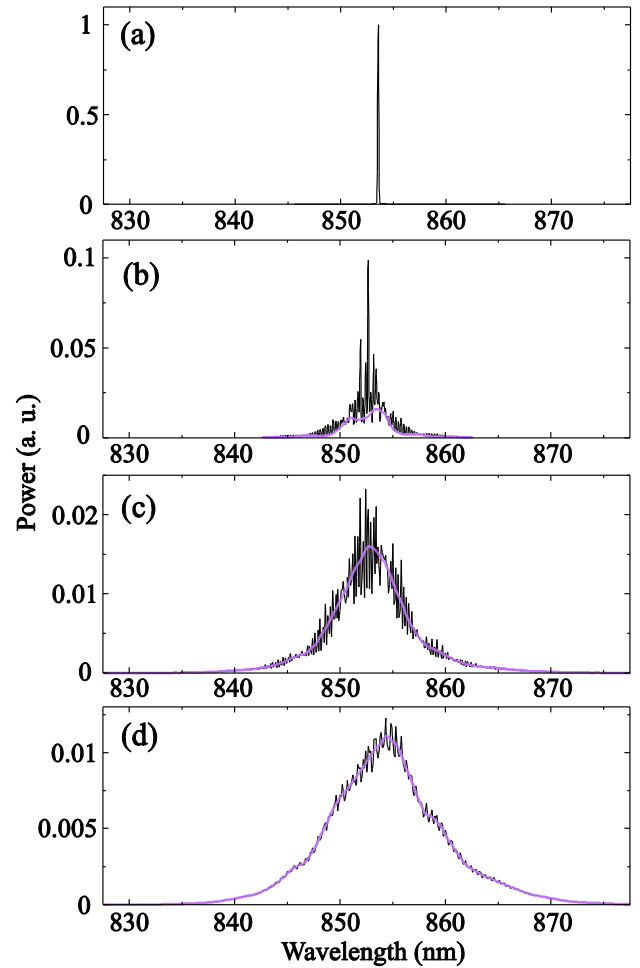

Fig. 3. The black lines show the spectra of the slave light with power $P_s$ = 51 mW for (a) master-light power $P_m$ = 0 (i.e., no master light injection), (b) $P_m$ = 0.3 mW, (c) $P_m$ = 0.8 mW and (d) $P_m$ = 2.7 mW. The pink (gray in the printed version) lines express smoothed curves for the broad component of the spectrum, $h_b(\lambda)$. Here, the smoothing in (b) was applied after removal of the laser mode components from the spectrum, while the raw spectra were smoothed for (c) and (d). The vertical axes are scaled so that the maximum power of the spectrum in (a) is 1.

To discriminate the spectrum under multimode oscillation from the noisy spectrum, the wavelength intervals between the spectrum's peaks and its temporal stability were examined. The peaks had nearly equal wavelength intervals under multimode oscillation, but were randomly scattered in the noisy spectrum. In addition, the spectrum was temporally stable under multimode oscillation but unstable when noisy-spectrum conditions were observed. With slave-light power $P_s$ = 15 and 51 mW, noisy spectra with unclear laser modes were recognized when $P_m \geq 0.2$ mW and $P_m \geq 0.4$ mW, respectively. Meanwhile, for $P_s$ = 150 mW, laser modes were observed up to the maximum master-light power $P_m = 2.7$ mW.

Here, the wavelength intervals of the spectrum peaks in Fig. 3(b) were as small as a third of the FSR. Thus, not all spectrum peaks corresponded to the longitudinal modes of the slave LD's cavity under multimode oscillation.

Three aspects of the slave light's spectrum were considered: the ratio of the power of incoherent light with respect to the overall power; the spectral linewidth of incoherent light; and spectral roughness caused by laser modes and noisy-component presence. To evaluate these aspects, a number of treatments were applied. First, if laser modes were observed in the raw spectrum $r(\lambda)$ ($\lambda$: wavelength), the broad component of the spectrum, $b(\lambda)$, was determined by removing the components of the laser modes from the raw spectrum. If not, $b(\lambda) = r(\lambda)$ was taken. Next, the curve for the broad component $b(\lambda)$ was smoothed using computer software via the application of a fast-Fourier-transform low-pass filter with a cutoff of 1 nm$^{-1}$. The smoothed curves $b_s(\lambda)$ are shown by the pink lines in Figs. 3 (b) – (d).

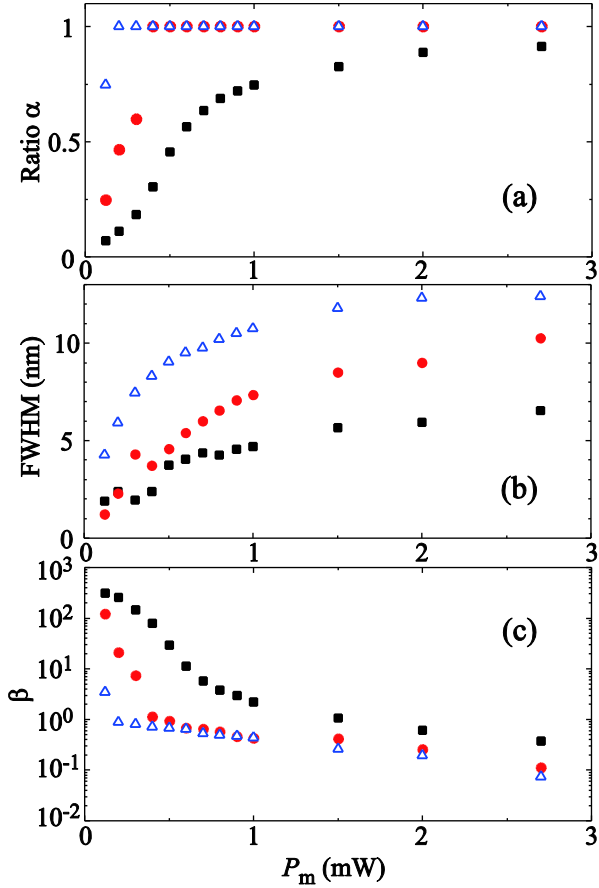

Fig. 4. (a) The $\alpha$ ratios (i.e., those of incoherent-light power to overall power), (b) the FWHMs of the incoherent-light spectra, and (c) the $\beta$ values representing spectral roughness as a function of master-light power $P_m$. Here, the triangles, circles and squares indicate cases with slave-light power $P_s = 15$, 51 and 150 mW, respectively.

The ratio of the incoherent light's power to the overall power was given by $\alpha = \int b_s(\lambda) d\lambda / \int r(\lambda) d\lambda$. Figure 4(a) shows the $\alpha$ ratios as a function of the master-light power $P_m$, where the triangles, circles and squares indicate cases of slave-light power $P_s = 15$, 51 and 150 mW, respectively. It can be seen that the ratios increased and became saturated when the master-light power rose. The saturation power (defined here as the master-light power $P_m$ at which the ratio $\alpha$ reached 0.5) increased with slave-light power, being < 0.1 mW at $P_s = 15$ mW and as high as 0.2 mW and 0.5 mW at $P_s = 51$ and 150 mW, respectively. As the ratio was 0.91 at the highest slave- and master-light power values (i.e., $P_s = 150$ mW and $P_m = 2.7$ mW), the power of incoherent light was estimated to be 140 mW.

The spectral linewidth of incoherent light was evaluated from the FWHM of $b_s(\lambda)$. Figure 4 (b) shows the FWHM values of $b_s(\lambda)$ as a function of master-light power $P_m$ for the three cases of slave-light power $P_s$. The linewidth basically increased with master-light power. When the slave-light power was lower, the FWHM at the same level of master-light power was larger in the region of $P_m \geq 0.3$ mW. At the maximum master-light power $P_m = 2.7$ mW, the FWHM values reached 12, 10 and 7 nm for $P_s = 15$, 51 and 150 mW, respectively.

To evaluate spectral roughness, the value $\beta = |r(\lambda_r) - b_s(\lambda_r)| / b_s(\lambda_r)$ was defined, where $\lambda_r$ is the wavelength at which $|r(\lambda) - b_s(\lambda)|$ was maximized. Figure 4(c) shows the values of $\beta$ as a function of the master-light power $P_m$ for the three cases of slave-light power $P_s$. It can be seen that $\beta$ decreased as master-light power increased, meaning that the slave light spectrum became smoother. The slope of the decrease in log$\beta$ was also more moderate when noisy spectra appeared ($P_m \geq 0.2$ mW at $P_s = 15$ mW, or $P_m \geq 0.4$ mW at $P_s = 51$ mW) than when laser modes were observed. In the case where a noisy spectrum appeared, the values of $\beta$ at $P_s = 15$ mW were comparable to those at $P_s = 51$ mW for identical levels of master-light power. Meanwhile, the values of $\beta$ at $P_s = 150$ mW were significantly larger than those at $P_s = 15$ and 51 mW for the same levels of master-light power. When the master-light power was increased to $P_m = 2.7$ mW, the values of $\beta$ fell to 0.07, 0.1 and 0.4 for $P_s = 15$, 51 and 150 mW, respectively.

## 3. Discussion

### A. Improvement to Achieve Higher Power and a Broader Spectrum

This section covers incoherent-light power enhancement based on the injection method. Because the slave-light power was equal to the output power from the LD without injection, an LD with higher output power can be expected to enable the achievement of incoherent light with a higher level of intensity. Although the master light must also be bright to enable the conversion of slave laser light with high power into incoherent light, a lack of master-light power may be overcome by injection in series using multiple LDs. As some commercially available cw LDs have an output power of several watts in the near-infrared region, the power of incoherent light may be enhanced to several tens of times higher than the level achieved in the experiment.

Next, let us consider the spectrum broadening of incoherent light based on the injection method. Under perfect phase locking, the slave light has the same spectral linewidth as the master light. In reality, however, the slave light had a significantly narrower spectrum. The most probable cause of this is limitation by the gain medium of the LD; it is unlikely that the spectral linewidth of the slave light significantly exceeds that of spontaneous emission from the medium. Here, radiation from the LD at an injection current below the threshold of laser oscillation has the spectrum of spontaneous emission from its medium. The FWHM of its spectrum was 10 nm, which corresponded relatively well with the maximum spectral linewidth of the slave light as determined in the experiment. Accordingly, it can be considered that the spectrum of the slave light may be further broadened by employing a slave LD that has a medium with a broader spontaneous-emission spectrum. Such LDs are difficult to purchase commercially because no broad spontaneous-emission spectrum is necessary in ordinary LD usage. However, considering the similarity between SLDs and LDs in terms of device structure (except for the cavity), it should be possible to improve the spectral linewidth of spontaneous emission from an LD's medium up to that of radiation from an SLD.

### B. Theoretical Consideration of the Principle

It may be assumed that incoherent light from the slave LD resulted from amplification of the master light due to stimulated emission in the slave LD's medium. Here, if this assumption were true, the slave-light power would increase with that of the master light. However, this was not the case in the experiment reported here; slave-light power was independent of master-light power. Accordingly, the experimental results support the authors' proposal that the slave light spectrum is broadened by master-light injection.

In several theories presented previously on injection locking, the stability of the phase lock was analyzed as a function of various parameters, such as the intensity of the master laser light and the Q factor of the slave laser's cavity [11, 16-20]. As it is considered that strong frequency modulation is applied to slave laser light outside the locking range [18, 19], the noisy spectra observed in the experiment may be explained qualitatively.

When master light is incoherent, analysis using the van der Pol equation or the Fokker-Planck equation is appropriate [11, 18]. For example, thermal-noise-related linewidth broadening of laser light has been calculated with noise regarded as an incoherent electromagnetic wave [18, 21]. Hence, it can be understood that the injection of incoherent light broadens the spectrum of slave light.

However, equations have been formulated to support the estimation of linewidth broadening assuming that slave light has clear laser modes. Accordingly, the experimental results in which the laser modes were broken by the injection of the master light may not be quantitatively comparable to the outcomes of conventional theoretical analysis. For example, let us apply the equation expressing linewidth broadening caused by thermal noise ($2\delta\nu \cong 4\pi h \nu (\Delta\nu_c)^2 n_T / P_s$) [18, 21] to the experimental case, where $h$, $\nu$, $\Delta\nu_c$ and $n_T$ are the Planck constant, the frequency of the slave light, the half-linewidth of the cavity resonance in the slave LD, and the photon number of thermal noise, respectively. If the photon number $n_T$ is replaced with that of the master light in the slave LD's cavity, $P'_m / (2\pi h \nu \Delta\nu_c)$, it can be considered that $2\delta\nu \cong 2(P'_m / P_s)\Delta\nu_c$, where $P'_m$ ($< P_m$) is the power of the master light passing through the facet of the slave LD. As the FSR is considerably larger than the linewidth of the cavity resonance, the equation does not explain the linewidth broadening that covers the FSR when $P_m/P_s << 1$.

Deeper theoretical analysis is necessary to explain the incomprehensible phenomenon observed in the experiment. In such analysis, the van der Pol equation or the Fokker-Planck equation will need to be solved in consideration of many slave light laser modes and a broad master-light spectrum. Additionally, the fact that the wavelength intervals between spectrum peaks under multimode oscillation were significantly smaller than the FSR of the slave LD's cavity may indicate that combination tones caused by the nonlinearity of the medium and sidebands induced by frequency modulation contribute to spectrum broadening.

## 4. Conclusion

In the study's experiment, the slave LD emitted incoherent light with a spectral linewidth of 7 nm and 140 mW of power based on the injection of incoherent light with 2.7 mW of power originating from the SLD. This method is expected to be important for the amplification of incoherent light, especially in wavelength ranges where taper amplifiers cannot be used. This optical system also provides a coherence-convertible light source.

The experimental results indicate spectrum broadening of the slave light based on the injection of incoherent master light rather than master light amplification caused by stimulated emission in the slave LD's medium. However, the phenomenon is difficult to explain in quantitative terms. To solve this problem, new theoretical analysis is required for a case in which many slave light laser modes are simultaneously affected by incoherent master light. The results of experimentation to clarify slave light spectrum characteristic variations caused by changes in master-light power are expected to be useful in the formulation of such analysis.

The authors are grateful to H. Inaba for allowing access to the optical spectrum analyzer used in the study.